\documentclass[a4paper,11pt]{article}
\usepackage[T1]{fontenc}
\usepackage[utf8]{inputenc}
\usepackage{amsfonts}
\usepackage{amssymb}
\usepackage{amsmath}
\usepackage{lmodern}
\usepackage{graphicx}
\usepackage[usenames,dvipsnames]{color}
\usepackage{graphicx}

\usepackage{float}
\newtheorem{Theorem}{Theorem}
\newtheorem{Lemma}{Lemma}
\newtheorem{Definition}{Definition}
\newtheorem{Proposition}{Proposition}

\title{Regularity of curve integrable spacetimes}
\author{Yafet Sanchez Sanchez\footnote{E-mail:Y.SanchezSanchez@soton.ac.uk} \\ 
Mathematical Sciences,\\ University of Southampton,\\ Southampton,\\ SO17 1BJ}

\begin{document}

\maketitle 

\begin{abstract}
 The idea of defining a gravitational singularity as
  an obstruction to
 the dynamical evolution of a test field
  (described by a PDE) rather
 than the dynamical evolution of a
  particle (described by a geodesics)
 is explored. In particular,
  the concept of field regularity is
 introduced which serves to
  describe the well-posedness of the local initial value problem for a
  given field. In particular this is applied to (classical) scalar
  fields in the class of curve integrable spacetimes to show that the
  classical singularities do not interrupt the well-posedness of the
  wave equation. 
\end{abstract}

\section{Introduction}

One of the biggest surprises that General Relativity (GR) has given us is
that under certain circumstances the theory predicts its own
limitations. There are two physical situations where we expect General
Relativity to break down. The first is the gravitational collapse of
certain massive stars when their nuclear fuel is spent. The second one
is the distant past of the universe when the density and temperature were
extreme. In both cases we expect that the geometry of spacetime will
show some pathological behaviour.
  
The first step towards a mathematical characterisation of the circumstances 
under which GR breaks down was achieved in the seminal work of
Penrose and Hawking in their singularity theorems (see \cite{hawking}, Chapter 8). 
The general structure of the theorems establish that
if on a spacetime (${M},g_{ab}$):
\begin{itemize}
  \item a condition on the curvature
  \item an appropriate initial or boundary condition
  \item and a global causal condition
  \end{itemize}
 are satisfied then (${M},g_{ab}$) must be
  geodesically incomplete \cite{senovilla}.
 
 The characterisation
  of singularities in terms of geodesic incompleteness requires us
  to consider spacetimes with metrics of differentiability at least
  $C^{1,1}$ (also denoted $C^{2-}$) in order to have a well-defined
  notion of geodesics. For any
 metric with lower differentiability
  one does not have uniqueness of
 the geodesic equation so it is not
  clear how one can define a precise
 notion of singular behaviour
  using point particles as probes to test
 the geometry. This was the
  motivation for Clarke to introduce the notion of a
 {\it $\square$-
    globally hyperbolic} spacetime \cite {generalized} which involves
  probing the geometry of spacetime with classical fields rather than
  point particles.

 Earlier work by Wald
    \cite{wald} gave a prescription to define dynamics in static
    non-globally hyperbolic spacetimes. Based on similar techniques,
    Kay and Studer \cite{kay} determined the boundary conditions for
    quantum scalar fields on singular spacetimes with conical
    singularities representing cosmic strings. Subsequently Horowitz and
    Marolf \cite{Marolf}, used Wald's approach to study the theory of
    quantum free particles in static spacetimes with timelike
    singularities. They used the term {\it quantum regular} if the
    evolution of any state is uniquely defined for all time. The main
    technique used for this was to notice that that if the the spatial
    Laplace-Beltrami operator is essentially self-adjoint in
    $L^{2}(\Sigma',\sqrt{h}d^{3}x)$ (where $\Sigma'$ is a three
    dimensional geodesically incomplete manifold and $\sqrt{h}d^{3}x$
    the volume form of the induced metric on $\Sigma'$) then using
    standard properties of self-adjoint operators, a unique evolution
    of the wave function is obtained. Moreover, the classical
    singularities disappear in the sense that there is no freedom in
    the boundary conditions to define the state evolution. Later work,
    by Ishibashi and Hasoya \cite{IH} used similar techniques to
    investigate the evolution of the wave equation $\square_{g}\phi=0$
    in static singular spacetimes by focusing on changing the
    function space from $L^{2}(\Sigma',\sqrt{h}d^{3}x)$ to
    $H^{1}(\Sigma',\sqrt{h}d^{3}x)$. The main reason for this is that
    finiteness of the energy states implies the finiteness of the $H^1$ norm. In
    addition, exploring different Hilbert spaces is important for
    quantum theory in curved spacetimes.
 Differences in the corresponding quantum field theories might provide
    useful insight into understand the behaviour of quantum states
    near singularities in curved spacetimes. Finally,  Ishibashi and Hasoya
    used the term {\it wave regular} if the initial value of the wave equation
    has unique solutions in the whole spacetime with no arbitrariness
    in the choice of boundary conditions. Vickers  and Wilson \cite{VW}  
    also studied the problem of conical singularities from Clarke's
    perspective and Wilson in \cite{conical} showed that one could 
    obtain dynamic evolution subject to constrains on the
    initial data and a flux condition in the singularity. 

    The link between the concept of $\square$- globally hyperbolic,
    quantum regular and wave regular is to redefine a singularity in
    spacetime not as an obstruction to geodesics or curves but as an
    obstruction to the dynamics of test fields. Nevertheless, each
    concept has it own characteristics. While quantum regularity
    probes spacetime with a quantum free particle, the notion of
    $\square$- globally hyperbolic and wave regular uses the classical
    wave equation. Also, while quantum regularity and wave regularity
    look to singularities in terms of boundary conditions, a
    $\square$-globally hyperbolic approach identifies the singularity
    as an interior point in a spacetime with low differentiability. In
    an heuristic manner, one can refer to {\it{field singularities}}
    as any approach to identify and characterise gravitational
    singularities as an obstruction to the evolution of test
    fields. This is in contrast to the standard approach where one
    uses geodesic incompleteness (which describes an obstruction to
    the evolution of a test particle) to identify singularities 

  In this paper the particular case we will  deal with is that of
  the wave equation and we define a point $p$ in $M$ to
  be {\it strongly wave regular} if there is a
    neighbourhood of $p$ in the spacetime ${({M},g_{ab})}$ such that
    there is a lens-shaped domain $\cal{U}$ containing $p$ (see figure
    1) and there is a triple $({\cal{P}},{\cal{Q}}, {\cal{R}}\times
    {\cal{S}})$ such that the
 initial value problem for the wave
    equation $\square_{g}\phi=f$ on ${\cal{U}}^{+}$ is locally well
    posed in the following sense:
\begin{itemize}
     \item There exists a solution in the function space ${\cal{P}}({\cal{U}}^{+},\sqrt{-g}d^{4}x)$.
     \item The solution is unique in the function space ${\cal{Q}}({\cal{U}}^{+},\sqrt{-g}d^{4}x)$.
     \item The solutions in the space ${\cal{Q}}({\cal{U}}^{+},\sqrt{-g}d^{4}x)$ depends continuously with respect initial data in function space ${\cal{R}}(\Sigma_{0},\sqrt{h}d^{3}x)\times{\cal{S}}(\Sigma_{0},\sqrt{h}d^{3}x)$.
   \end{itemize}

A {\it weakly wave regular} point $p$ only satisfies the first two conditions. A {\it strongly  wave regular spacetime} is defined to be one such that every point $p$ in ${({\cal{M}}, g_{ab})}$ is strongly wave regular.

 In
this paper we refine and provide full details of Clarke's arguments
in
 \cite{generalized} where he outlined this issue for what he
called
 curve integrable spacetimes and give a detailed description
of the
 techniques needed to show that solutions to the wave
equations
 exist and are unique. Roughly speaking a curve
integrable spacetime  is one in which the integrals of both the
connection and curvature along a curve are bounded (see condition
(4) of the main theorem for a precise description). From a physical
point of view a  spacetime is a curve integrable spacetime if there is
a set $C$ that defines a range of timelike directions which are
transverse to any shock or caustic that may be present. The theorem
proved by Clarke in \cite{generalized} required both the quadratic
and linear part of the Riemann tensor (in terms of the Christoffel
symbols) to be separately integrable along the timelike
directions. However Clarke also conjectured that one can prove
existence and uniqueness of the wave equation if one only required
the weaker condition of the integrability of the Christoffel symbols
and the Riemann tensor. In the present paper we prove this
conjecture. In addition, we also show continuity
with respect initial data and establish strong wave regularity.

  \section{Curve Integrable Spacetimes}

  In this section we establish the well-posedness of solutions to the
  initial value problem for $\square_{g}\phi=f$ where $\square_{g}$ is
  the wave operator given by a metric $g_{ab}$ that corresponds to a
  curve integrable spacetime. The proof is in four main steps. The
  first step is to define an energy inequality which is an inequality
  between an integral of the function and its derivatives at a future
  time bounded above by an integral of the function and its
  derivatives at the initial time and the source function. The second
  step is to show self-adjointness of the wave operator in some
  appropriate function space and to use the Hahn-Banach theorem to prove
  existence of a solution. The third step is to notice that the energy
  inequality allows us to show uniqueness of our solution. The last
  step is to conclude that the solution depends continuously on the
  initial data again using the energy inequality.

  Following Clarke \cite{generalized}, we introduce an enlarged notion
  of a solution for these spacetimes. The general geometric background
  we use to define a generalised solution is the existence of a
  \emph{lens-shaped} domain $\cal{U}$ properly contained in a open subset
  $\Omega$ of a $4$-dimensional spacetime $({M},g_{ab})$.

This means that there is a smooth map $\Theta :\Sigma\times
(-a,a)\rightarrow {M}$ where
$\Sigma\subset{M}$ is a compact, $C^{1}$ co-dimension 1
sub-manifold with boundary, with the property that:
\begin{itemize}
  \item$\Theta (\cdotp ,0):\Sigma\rightarrow \Sigma_{0}$ is the identity map.
   \item $\Theta (x,\tau)=\Theta (x,s)$ for any $x\in\partial \Sigma$, $\tau,s\in(a,b)$.
  \item  for any fixed $s\in(a,b)$, $\Theta (S,s)$ is an  $3$- dimensional spacelike hypersurface.
  \item  away from $\partial \Sigma\times(a,b)$, $\Theta $ is a diffeomorphism.\\
\end{itemize}
We denote the region from $0$ to $\tau\ge0$ by ${\cal{U}}_{\tau}^{+}$
and from $\tau'\le0$ to $0$ by ${\cal{U}}_{\tau'}^{-}$. Notice that
given coordinates $x^i$ on $\Sigma$, $\Theta $ provides coordinates
$(\tau,x^{i})$ for the region ${\cal{U}}_{\tau}^{+}$ away from the image
of $\partial \Sigma$. We will therefore always choose charts such that
the time coordinate coincides with the time coordinate given by
$\Theta$ when working in coordinates.

\begin{figure}[H]
\centering
\includegraphics[width=90mm]{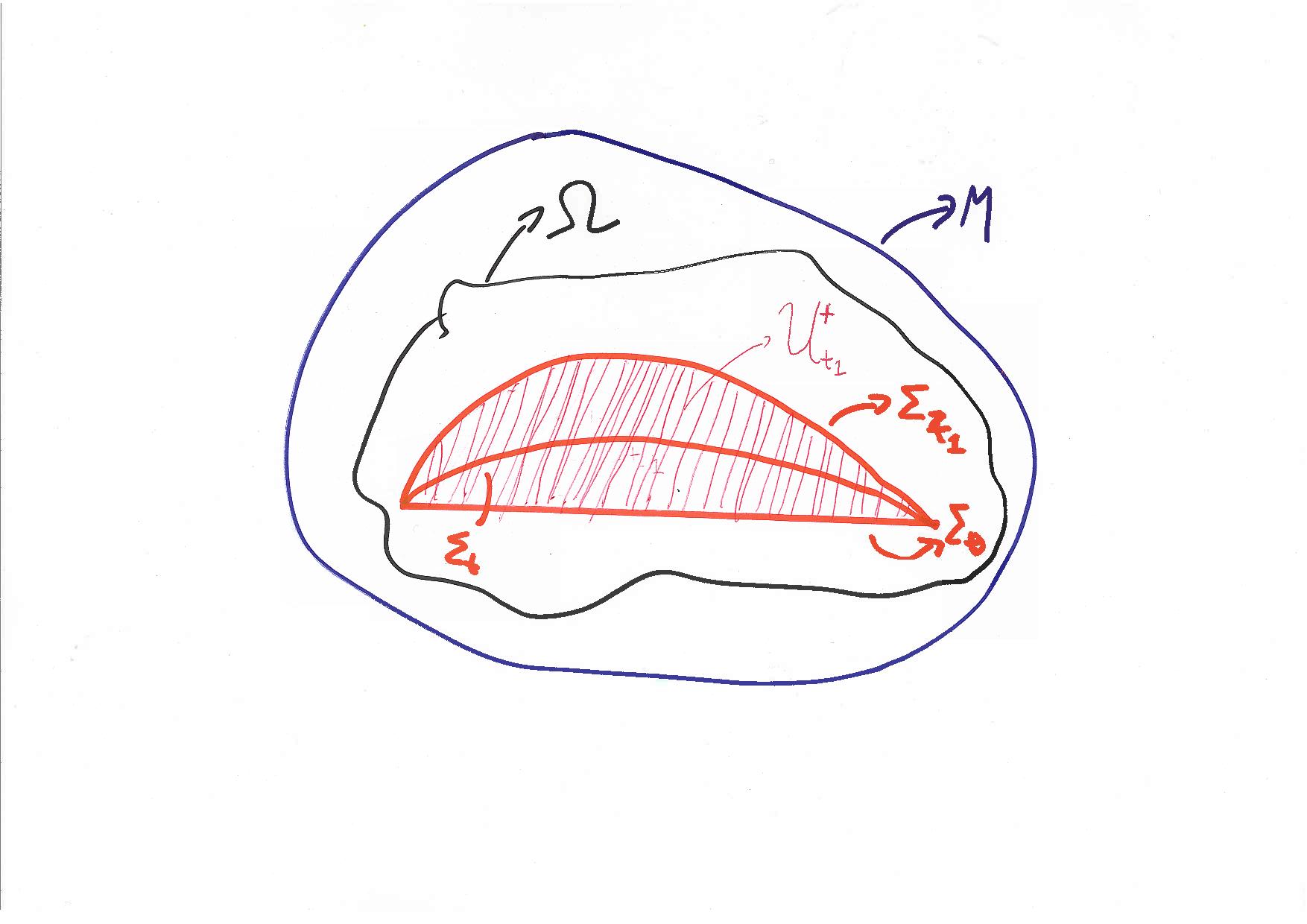}
\caption{The general geometric setting.}
\label{Lens-shaped domain}
\end{figure}

We start by considering the case of  a $C^{2}$ solution to the problem:
 \begin{equation}\label{wp1}
          \square_{g}\phi=f
  \end{equation}
  in the region ${\cal{U}}_{\tau}^{+}$ where $g_{ab}$ is a smooth metric with $C^{2}\times C^{1}$ initial data $(\varphi,\pi)$
    
    \begin{equation}\label{wp2}
         \phi|_{\Sigma_{0}}=\varphi  
    \end{equation}
         
         \begin{equation}\label{wp3}
     \frac{\partial \phi }{\partial t}|_{\Sigma_{0}}= \pi    
         \end{equation}
  
 Then  multiplying \eqref{wp1} by a test field $\rho\in C_{o}^{\infty}(\cal{{\cal{U}}_{\tau}^{+}})$ and integrating by parts gives:

\begin{eqnarray}\label{generalisedsol}
        \int_{{\cal{U}}_{\tau}^{+}}\frac{\partial\phi}{\partial x^{i}}(\frac{\partial\rho}{\partial x^{j}} g^{ij}) \sqrt{-g}d^{4}x &=&- \int_{{\cal{U}}_{\tau}^{+}}\rho f\sqrt{-g}d^{4}x-\int_{\Sigma_{0}}\rho\pi \sqrt{h}d^{3}x\\\label{generalisedsol1}
    \phi\arrowvert _{\Sigma_{0}}&=&\varphi 
 \end{eqnarray}

 Clarke then generalised the idea of a solution to
 (\ref{wp1}),(\ref{wp2}) and (\ref{wp3}) in the region
 ${\cal{U}}_{\tau}^{+}$ for a low differentiable metric to be a
 function $\phi$ that satisfies condition (\ref{generalisedsol})\ $
 \forall \rho \in {\cal{D}}({\cal{{\cal{U}}_{\tau}^{+}}})$ with
 initial value (\ref{generalisedsol1}). Notice that the expression is
 well defined for $\phi\in H^{1}({\cal {U}}_{\tau}^{+},
 \sqrt{-g}d^{4}x)$, $\varphi\in H^{1}(\Sigma_{0}, \sqrt{h}d^{3}x)$,
 $\pi\in L^{2}(\Sigma_{0}, \sqrt{h}d^{3}x)$, $f\in L^{2}({\cal
   {U}}_{\tau}^{+}, \sqrt{-g}d^{4}x)$ and $g^{ij}\in L^{2}({\cal
   {U}}_{\tau}^{+}, \sqrt{-g}d^{4}x)$ where $\sqrt{-g}d^{4}x$ is the
 volume element of the metric $g_{ab}$ and $\sqrt{h}d^{3}x$ is the
 induced volume given by the induced metric $h_{ab}$ on $\Sigma_{0}$.
 To simplify notation we define $\nu_{g}:=\sqrt{-g}d^{4}x$ and $\mu_{h}:=\sqrt{h}d^{3}x$.

\subsection{The main theorem}

The theorem we will prove can be stated precisely as:

\begin{Theorem}\label{main1}
Let ($M,g$) be a Lorentzian manifold and $p$ a point in an open subset $\Omega\subset M$ with compact closure such that there is a lens-shaped domain ${\cal{U}}$ satisfying:

\begin{enumerate}
  \item The components $g_{ij}$ and $g_{ij}^{-1}$ are $C^{0}$ ;
  \item The components $g_{ij}$ are $C^{1}$ in $M\backslash J^{+}(p)$
  \item $g_{ij,k}$ exist and are $L_{loc}^{p}(M)$ with $p\ge2$;
  \item there exist functions $R^{i}_{jkl}$ which, interpreted as distributions, coincide with the Riemann tensor defined distributionally from $g$ and $g_{ij,k}$. 
  \item there is a non-empty set, $C\subset\mathbb{R} ^{4}$, and positive functions $ M,N:\mathbb{R} ^{+}
    \rightarrow \mathbb{R} ^{+}$ such that, if $\gamma$ is a curve with $d\gamma/ds\in C$ for all s then
  
  \begin{itemize}
    \item $\gamma$ is future timelike;
    \item the integrals
    \begin{equation}
      I_{\gamma}(a):=\int^{a}_{0}|\Gamma^{i}_{jk}(\gamma(s))|ds
    \end{equation}
     and
     \begin{equation}
      J_{\gamma}(a):=\int^{a}_{0}|R^{i}_{jkl}(\gamma(s))|ds
     \end{equation}
  (where $\Gamma$ is defined using the weak derivatives of $g$) are convergent, with
  \begin{equation}
    I_{\gamma}(a)<M(a),  J_{\gamma}(a)<N(a)
  \end{equation}
  and $M(a), N(a) \rightarrow 0, \mbox{as a} \rightarrow 0$.
 \end{itemize}
 \end{enumerate} 
 
then  $p$ is strongly wave regular.

\end{Theorem}

\subsection{Energy inequality}
We follow the classical method of proving well posedness of the
wave equation by using an energy inequality as shown for example in
Hawking and Ellis \cite{hawking}, Clarke \cite{generalized} and Wilson
\cite{conical}. The energy inequality gives an integral of the
function and its derivatives at a future time bounded above by an
integral of the function and its derivatives at the initial time and
an integral of the source function over the region between the initial
time and the future time. We will first assume that $g_{ab}$ is smooth.
Then at the end we will give the extra requirements that
the metric must satisfy in order that the energy inequality is still
valid when the differentiability is below $C^{2}$.

Consider $\phi$ a solution to the wave equation that is $C^{2}$ with
energy-momentum tensor $T^{ab}$ given by:
\begin{equation}
       T^{ab} = \left(g^{ac}g^{bd}-\frac{1}{2}g^{ab}g^{cd}\right)\frac{\partial\phi}{\partial x^{c}}\frac{\partial\phi}{\partial x^{d}}-\frac{1}{2}g^{ab}\phi^{2}\label{energytensor}
\end{equation}

 Now choosing a smooth timelike vector field,  $\Upsilon ^{a}$, (which in the theorem will be chosen to be the $4$-velocity of a family of observers) we define the energy integral:
 
 \begin{equation}
 E(\tau)=\int_{\Sigma_{\tau}}T^{ab}\Upsilon_{a}n_{b}\mu_{h}
\end{equation}
where $n^{a}$ is a future pointing vector normal to  $\Sigma_{\tau}$.

Then we use the divergence theorem on the domain $\overline{\cal{U}_{\tau}^{+}}$:

\begin{equation}\label{stokess}
  \int_{{{\cal{U}_{\tau}^{+}}}}\mbox{div}\left(T^{ab} \Upsilon_{a}\right)\nu_{g}=\int_{\partial{{\cal{U}_{\tau}^{+}}}}T^{ab}\Upsilon_{a}n_{b}\mu_{h}
\end{equation}

The left hand side takes the explicit form:
\begin{equation}\label{left0}
    \int_{{\cal{U}_{\tau}^{+}}}\left(g^{ab}\frac{\partial\phi}{\partial x^{b}}\Upsilon_{a}\right)\left[f-\phi\right]+T^{ab}\nabla_{b}\Upsilon_{a}N\sqrt{h}d^{4}x
      \end{equation} 
where $f=\square_{g}\phi$, $N$ is the lapse function which satisfies $N dt=n_{a}$ and $\sqrt{h}$ is the induced metric in the hypersurface $\Sigma_{0}$.

The right hand side then takes the form:
\begin{equation}
\left(\int_{\Sigma_{\tau}}-\int_{\Sigma_{0}}\right)T^{ab}\Upsilon_{a}n_{b}\mu_{h}.
\end{equation}

Now we introduce the following norms:

\begin{equation}
\tilde{\Arrowvert} \phi\Arrowvert^{1}_{\Sigma_{\tau}}=\left[\int_{\Sigma_{\tau}}\left(\frac{\partial \phi}{\partial \tau}\right)^{2}+\sum^{3}_{i=1}\left(\frac{\partial \phi}{\partial x^{i}}\right)^{2}+\phi^{2} \mu_{h}\right]^{\frac{1}{2}}
\end{equation}

\begin{equation}
  \Arrowvert\phi\Arrowvert^{1}_{{{\cal{U}_{\tau}^{+}}}}=\left[\int_{{{\cal{U}_{\tau}^{+}}}}\left(\frac{\partial \phi}{\partial \tau}\right)^{2}+\sum^{3}_{i=1}\left(\frac{\partial \phi}{\partial x^{i}}\right)^{2}+\phi^{2} \nu_{g}\right]^{\frac{1}{2}}
\end{equation}

\begin{equation}
  \Arrowvert\phi\Arrowvert^{0}_{{{\cal{U}_{\tau}^{+}}}}=\left[\int_{{{\cal{U}_{\tau}^{+}}}}\phi^{2}\nu_{g}\right]^{\frac{1}{2}}
\end{equation}

The main reason for introducing these norms which are naturally related to Sobolev type norms is the following result which can be found in \cite{conical}.
\begin{equation}\label{wilson11}
C_{1}E(\tau)\le(\tilde{\Arrowvert} \phi\Arrowvert^{1}_{\Sigma_{\tau}})^{2}\le C_{2} E(\tau)
\end{equation}
for constants $C_{1},C_{2}\ge0$.

Also notice that:

\begin{equation}\label{wilson22}
  \Arrowvert\phi\Arrowvert^{1}_{{{\cal{U}_{\tau}^{+}}}}\le k \left(\int^{\tau}\left(\tilde{\Arrowvert} \phi\Arrowvert^{1}_{\Sigma_{t}}\right)^{2}dt\right)^{\frac{1}{2}}
\end{equation}
where $k$ is a constant that depends on $g_{ab}$.

Now we can obtain the following bounds for all the terms in (\ref{left0}):

  \begin{eqnarray}
    \int_{0}^{\tau}\left(\int_{\Sigma_{t}}\left(g^{ij}\frac{\partial\phi}{\partial x^{j}}\Upsilon _{i}\right)f N\mu_{h}\right)dt\le K_{1}( (\Arrowvert\phi\Arrowvert^{1}_{U_{t}})^{2}+ (\Arrowvert f\Arrowvert^{0}_{U_{t}})^{2})\label{l1}
  \end{eqnarray}

  \begin{eqnarray}
    \int^{\tau}_{0}\left(\int_{\Sigma_{t}}\left(g^{ij}\frac{\partial\phi}{\partial x^{j}}\Upsilon_{i}\right)\phi N\mu_{h}\right)dt\le K_{2}(\Arrowvert\phi\Arrowvert^{1}_{U_{t}})^{2}\label{l2}
     \end{eqnarray}
 
  where $K_{1},K_{2}$ are constants that depend on $g_{ab}$ and $\Upsilon^{a}$.

  The last term in (\ref{left0}) is bounded by:
 
  \begin{equation}\label{r1}
    \int_{{\cal{U}_{\tau}^{+}}}T^{ij}\nabla_{j}\Upsilon_{i}\le K_{3} (\Arrowvert \phi\Arrowvert_{{\cal{U}_{\tau}^{+}}}^{1})^{2}
  \end{equation}
  where $K_{3}$ is a constant that depend on $g_{ab}$ and $\nabla_{b}\Upsilon_{a}$.

  Estimating all the terms in (\ref{stokess}) by the bounds available (\ref{l1}),(\ref{l2}) and (\ref{r1})  gives the inequality:
  
  \begin{equation}\label{preeee}
    E(\tau)\le E(0) + k_{0} (\Arrowvert f\Arrowvert_{{\cal{U}_{\tau}^{+}}}^{0})^{2} + k_{1}(\Arrowvert \phi\Arrowvert_{{\cal{U}_{\tau}^{+}}}^{1})^{2}
  \end{equation}
  where $k_{0},k_{1}$ are positive constants that depend on the metric $g_{ab}$, the vector field $\Upsilon_{a}$ and the covariant derivative $\nabla_{b}\Upsilon_{a}$.
    
 Now rewriting (\ref{preeee}) using (\ref{wilson11}) and (\ref{wilson22}) as : 
  
  \begin{equation}
   E(\tau)\le E(0) + k_{0}(\Arrowvert f\Arrowvert_{{\cal{U}_{\tau}^{+}}}^{0})^{2} +  k_{2}\int^{\tau}E(\tau')d\tau'
  \end{equation} 
  
  Using Gronwall's inequality the desired energy inequality is obtained:
  
  \begin{eqnarray}
     E(\tau) &\le&K_{4} (E(0)+ (\Arrowvert f\Arrowvert_{{\cal{U}_{\tau}^{+}}}^{0})^{2})\label{ee1}
              \text{ for all } \tau\le t_{1}
  \end{eqnarray}
 where $K_{4}$ positive constant that depends on the chosen finite time $t_{1}$,the metric $g_{ab}$, the vector field $\Upsilon_{a}$ and the covariant derivative $\nabla_{b}\Upsilon_{a}$.
 
  In term of the Sobolev norms we obtain the expression:  
\begin{equation}\label{dasenergy}
( \tilde{\|} \phi\|^{1}_{\Sigma_{\tau}})^{2} \lesssim  (( \tilde{\|} \phi\|^{1}_{\Sigma_{0}})^{2} + (\Arrowvert f\Arrowvert_{{\cal{U}_{\tau}^{+}}}^{0})^{2})
  \end{equation} 
  where we say $a\lesssim b$ if $a\le kb$ for some constant $k$.
   
  We have proved all these results for $C^{2}$ functions but we can readily 
  extend them to the subspace of solutions of (\ref{wp1}) in  $H^{1}({\cal {U}}_{\tau}^{+},
  \nu_{g})$ with all derivatives in $L^{2}(\Sigma_{\tau},\mu_{h})$  for all
  $\tau\in[0,t_{1}]$.

We now look for the conditions required to obtain again the energy inequalities in the low differentiable setting. The basic requirement we need is that the we can apply Stokes' Theorem. To our knowledge the optimum results are given by the following theorem:

\begin{Theorem}\label{cstoke}
Let $\Omega$ be a compact set with compact closure  with Lipschitz boundary  and let $Z^{a}$ be a vector field on an $(n+1)$- dimensional manifold
$M$ with continuous metric $g_{ab}$ and metric volume element $\nu_{g}$. If
$$Z^{a}\in W^{1,1}_{loc}({M}), g_{ab} \in W^{1,n+1}_{loc}(M),$$
then the Stokes identity holds:
$$\int_{\partial \Omega}j=\int_{\Omega}dj$$
for $j=i_{Z^{a}}\nu_{g}$ and where $dj=div(Z^{a})\nu_{g}$
\end{Theorem}
A proof of this can be found in \cite{crush}.

Unfortunately, the result can not be applied directly to our case to obtain the
energy inequalities as it would require us to take $\phi\in
H_{loc}^{2}(M)$. The way to proceed is to again first assume
$\phi\in C^{2}$ and look for the conditions needed on the metric and
the vector field $\Upsilon_{a}$ such that we obtain again the energy
inequalities.

Assume now that the metric and its metric volume element are
continuous. This is enough to satisfy the hypothesis and allows us to apply Stokes' theorem. Of course we would like to have the same expression as in (\ref{left0}). This requires
the existence of a metric connection, i.e. $\nabla g_{ab}=0$. In
\cite{roland}, it is stated that sufficient conditions for the
existence of a Levi-Civita connection are the existence of a
connection $\nabla\in L_{loc}^{2}({M},\nu_{g})$ and that
$g^{ab}\in L_{loc}^{\infty}(M)$. We say $\nabla\in
L_{loc}^{2}({M},\nu_{g})$ if $\nabla_{X_{a}}Y^{a}\in
L_{loc}^{2}({M},\nu_{g})$ for any pair $X_{a},Y_{a}$ of
$C^{\infty}$ vector fields. If the Christoffel symbols satisfy
$\Gamma^{i}_{jk}\in L_{loc}^{2}({M},\nu_{g})$ then they define
a Levi-Civita connection. For example, the class of Geroch-Traschen
class of metrics satisfy the above conditions.

It can be seen by direct inspection that the other inequalities require
only that $g^{ab}\in L_{loc}^{\infty}(M)$ and $g_{ab}\in
L_{loc}^{\infty}(M)$ which are also enough to maintain the
results (\ref {wilson11}) and (\ref{wilson22}). Finally the conditions
on the timelike vector field $\Upsilon_{a}$ which is needed is that
the covariant derivative is essentially bounded. The sufficient
analytical conditions which guarantee the existence of such vector field are
a key part of our result and are established in Proposition \ref{append} shown in the appendix.

Now we have all the analytical conditions needed to recover the energy
inequality (\ref{dasenergy}) we can again extend the result to the
subspace of solutions of (\ref{wp1}) in $H^{1}({\cal {U}}_{\tau}^{+}, \nu_{g})$ with all derivatives in $L^{2}(\Sigma_{\tau},\mu_{h})$  for all
  $\tau\in[0,t_{1}]$. 

For clarity we state the result as a Lemma:
\begin{Lemma}\label{main}
  If ($M,g$) is a Lorentzian manifold and there is subset
  $\Omega$ with compact closure such that there is a lens-shaped
  domain, ${\cal {U}_{\tau}^{+}}$, in $\Omega$ satisfying:

\begin{enumerate}
  \item The components $g_{ij}$ and $g_{ij}^{-1}$ are $C^{0}$ ;
  \item The components $g_{ij}$ are $C^{1}$ in $M\backslash J^{+}(p)$;
  \item weak derivatives $g_{ij,k}$ exist and are $L_{loc}^{2}({M},\nu_{g})$;
  \item there exist functions $R^{i}_{jkl}$ which, interpreted as distributions, coincide with the Riemann tensor defined distributionally from $g$ and $g_{ij,k}$. 
  \item there is a non-empty set, $C\subset\mathbb{R} ^{4}$, and positive functions $ M,N:\mathbb{R} ^{+}
    \rightarrow \mathbb{R} ^{+}$ such that, if $\gamma$ is a curve with $d\gamma/ds\in C$ for all s then
  
  \begin{itemize}
    \item $\gamma$ is future timelike;
    \item the integrals
    \begin{equation}
      I_{\gamma}(a):=\int^{a}_{0}|\Gamma^{i}_{jk}(\gamma(s))|ds
    \end{equation}
     and
     \begin{equation}
      J_{\gamma}(a):=\int^{a}_{0}|R^{i}_{jkl}(\gamma(s))|ds
     \end{equation}
  (where $\Gamma$ is defined using the weak derivatives of $g$) are convergent, with
  \begin{equation}
    I_{\gamma}(a)<M(a),  J_{\gamma}(a)<N(a)
  \end{equation}
  and $M(a), N(a) \rightarrow 0, \mbox{as a} \rightarrow 0$.
 \end{itemize}
 \end{enumerate} 
then 

\begin{equation}
( \tilde{\|} \phi\|^{1}_{\Sigma_{\tau}})^{2} \lesssim  (( \tilde{\|} \phi\|^{1}_{\Sigma_{0}})^{2} + (\Arrowvert f\Arrowvert_{{\cal{U}_{\tau}^{+}}}^{0})^{2})
  \end{equation} 
for all $\phi\in H^{1}({\cal {U}}_{t_{1}}^{+}, \nu_{g})$  with all derivatives in $L^{2}(\Sigma_{\tau},\mu_{h})$  for all  $\tau\in[0,t_{1}]$ and $\square_{g}\phi=f$ .
\end{Lemma}

\subsection{Self-adjointness and Existence}

The next step required to establish the theorem is the self-adjointness of the operator $\square_{g}$ in an appropriate function space.  

Consider the $L^{2}({\cal {U}}_{t_{1}}^{+},\nu_{g})$ norm.  

\begin{equation}
 \langle \psi,\omega\rangle=\int_{{{\cal{U}}_{t_{1}}^{+}}}\psi\omega \nu_{g} 
\end{equation}

We shall single out two important subspaces of $H^{1}({\cal{U}}_{t_{1}}^{+},\nu_{g})$:

\begin{eqnarray}
    V_{\Sigma_{0}}&=&\{\psi\in C^{\infty}(\Omega)  \mbox{ s. t. } \psi|_{\Sigma_{0}}=n^{a}\psi,_{a}|_{\Sigma_{0}}=0 \text{ and } \square_{g}\psi \in L^{2}({\cal {U}}_{t_{1}}^{+},\nu_{g}) \}\nonumber\\
  V_{\Sigma_{t_{1}}}&=&\{\omega\in C^{\infty}(\Omega)\mbox{ s. t. } \omega|_{\Sigma_{t_{1}}}=n^{a}\omega,_{a}|_{\Sigma_{\tau}}= 0 \text{ and } \square_{g}\psi \in L^{2}({\cal {U}}_{t_{1}}^{+},\nu_{g}) \}\nonumber
\end{eqnarray}

Then the condition that $g_{ij}$ is $C^{0}$ and that $Z_{a}=\omega\psi_{a}\in C^{\infty}(\Omega)$ and $Z_{a}=\psi\omega_{a}\in C^{\infty}(\Omega)$ is enough to apply Theorem \ref{cstoke} and obtain:

\begin{eqnarray}
\int_{{{\cal{U}}_{t_{1}}^{+}}} g^{ab}\psi_{b}\omega_{a} \nu_{g} + \int_{{{\cal{U}}_{t_{1}}^{+}}} \omega\square_{g}\psi \nu_{g}&=&0\\
\int_{{{\cal{U}}_{t_{1}}^{+}}} g^{ab}\psi_{b}\omega_{a} \nu_{g} + \int_{{{\cal{U}}_{t_{1}}^{+}}} \psi\square_{g}\omega \nu_{g}&=&0
\end{eqnarray}\\

so combining the above equations give
\begin{eqnarray}
\int_{{{\cal{U}}_{t_{1}}^{+}}} \square_{g}\psi\omega \nu_{g}
&=&
\int_{{{\cal{U}}_{t_{1}}^{+}}} \psi\square_{g}\omega \nu_{g}
\end{eqnarray}

and we obtain

\begin{equation}
  \langle\square_{g}\psi,\omega\rangle=\langle\psi,\square_{g}\omega\rangle
\end{equation}

which proves the self-adjointness of $\square_{g}$ for $\psi\in V_{\Sigma_{0}},\omega\in V_{\Sigma_{t_{1}}}$.

In the following lemma we establish a density result needed later .

\begin{Lemma}
The spaces $V_{\Sigma_{0}},V_{\Sigma_{t_{1}}}$ are dense in $L^{2}({\cal {U}}_{t_{1}}^{+},\nu_{g})$.
\end{Lemma}

 {\it{Proof of Lemma. }}It is enough to prove that $ C^{\infty}_{o}({\cal{U}}_{t_{1}}^{+})$ is contained in both sets.\cite{pde}
  
 Let $\phi\in C^{\infty}_{o}({\cal{U}}_{t_{1}}^{+})$ The only thing we
 need to prove is that $\square_{g}\phi\in
 L^{2}({\cal{U}}_{t_{1}}^{+},\nu_{g})$. Explicitly in coordinates we
 have that $$\square_{g}\phi=g^{ij}\frac{\partial\phi}{\partial
   x^{i}\partial x^{j}}+\Gamma^{k}_{ij}\frac{\partial\phi}{\partial
   x^{k}}.$$ That the first term is in
 $L^{2}({\cal{U}}_{t_{1}}^{+},\nu_{g})$ follows directly from $g^{ij}
 \in C^{0}$ and $\phi\in C^{\infty}_{o}({\cal{U}}_{t_{1}}^{+})$. The
 second term is in $L^{2}({\cal{U}}_{t_{1}}^{+},\nu_{g})$ because
  
  \begin{eqnarray}
    \int_{{\cal{U}}_{t_{1}}^{+}}(\Gamma^{k}_{ij}\frac{\partial\phi}{\partial x^{k}})^{2}\nu_{g}&\le&\left\Arrowvert\left(\frac{\partial\phi}{\partial x^{k}}\right)^{2}\right\Arrowvert_{\infty} \int_{{\cal{U}}_{t_{1}}^{+}}\left(\Gamma^{k}_{ij}\right)^{2}\nu_{g}\\
    &<&\infty
  \end{eqnarray}  
  where we have used the fact that the connection is in
  $L_{loc}^{2}({M},\nu_{g})$ and Holder's inequality.  This
  allows us to conclude then that $\square_{g}\phi\in
  L^{2}({\cal{U}}_{t_{1}}^{+})$ hence $\phi$ is in
  $V_{\Sigma_{0}}, V_{\Sigma_{t_{1}}}$ for any $\phi\in C^{\infty}_{o}({\cal{U}}_{t_{1}}^{+})$. $\boxdot $\\

  The proof of existence uses the Hahn-Banach Theorem and the Riesz
  Representation Theorem (see \cite{pde} p 199).
 We start by proving that $\square_{g} V_{\Sigma_{0}}$ is dense in
  $L^{2}({\cal {U}}_{t_{1}}^{+},\nu_{g})$. It is enough to prove
  that $$\square_{g} V_{\Sigma_{0}}^{\bot}=\{x\in L^{2}({\cal
    {U}}_{t_{1}}^{+},\nu_{g}) | \langle x,y\rangle \text{ for all }
  y\in\square_{g} V_{\Sigma_{0}}\}=0.$$

  Suppose $\eta \in\square_{g} V_{\Sigma_{0}}^{\bot}$, because
  $V_{\Sigma_{t_{1}}}$ is dense in
  $L^{2}({\cal{U}}_{t_{1}}^{+},\nu_{g})$, there exist a sequence
  $\{\omega_{n}\}$ in $V_{\Sigma_{t_{1}}}$ converging to $\eta $ in
  $L^{2}({\cal{U}}_{t_{1}}^{+},\nu_{g})$.

Now we have:

\begin{equation}
  \langle\phi,\square_{g}\omega_{n}\rangle=\langle\square_{g}\phi,\omega_{n}\rangle\rightarrow\langle\square_{g}\phi,\eta \rangle=0
\end{equation}  
for all $\phi\in V_{\Sigma_{0}}$. Since, $V_{\Sigma_{0}}$ is dense, this implies that  $(\Arrowvert \square_{g}\omega_{n}\Arrowvert_{{\cal{U}_{\tau}^{+}}}^{0})^{2}\rightarrow 0$. 

In order to continue the proof we need the result of the following Lemma.

\begin{Lemma}

 For any element $\xi$ in $\Sigma_{0}$ we have the following form energy inequality:
\begin{equation}
( \tilde{\|} \xi\|^{1}_{\Sigma_{\tau}})^{2} \lesssim   (\Arrowvert \square_{g}\xi\Arrowvert_{{\cal{U}_{\tau}^{+}}}^{0})^{2}
  \end{equation} 
 
 for $0\le\tau\le t_{1}$.

\end{Lemma}

{\it{Proof of the Lemma}}.  
We prove the energy inequality by choosing
a future pointing normal vector field. However, we can also repeat the
argument in an identical form by choosing a past-pointing normal
vector field to $\Sigma_{t_{1}}$

In that case we obtain the energy inequality: 

\begin{equation}\label{v1}
( \tilde{\|} \phi\|^{1}_{\Sigma_{\tau}})^{2} \lesssim  (( \tilde{\|} \phi\|^{1}_{\Sigma_{t_{1}}})^{2} + (\Arrowvert \square_{g}\phi \Arrowvert_{{\cal{U}_{\tau}^{+}}}^{0})^{2})
  \end{equation} 

 for $0\le\tau\le t_{1}$.
 
  Then energy inequality takes the form: 
 
 \begin{equation}\label{v2}
( \tilde{\|} \omega\|^{1}_{\Sigma_{\tau}})^{2} \lesssim   (\Arrowvert \square_{g}\omega\Arrowvert_{{\cal{U}_{\tau}^{+}}}^{0})^{2}
  \end{equation} 
 for $0\le\tau\le t_{1}$ and $\omega\in V_{\Sigma_{t_{1}}}$. $\boxdot $ \\

Using now the energy inequality (\ref{v1}) with (\ref{wilson22}) and integrating both sides from $0$ to $t_{1}$ we obtain:

\begin{equation}\label{ipp}
(\Arrowvert \omega_{n}\Arrowvert^{1}_{{\cal {U}}_{t_{1}}^{+}})^{2}\lesssim \int_{0}^{t_{1}} (\Arrowvert \square_{g}\omega_{n}\Arrowvert_{{\cal{U}}_{\tau}^{+}}^{0})^{2}d\tau
\end{equation}

Hence, $(\Arrowvert \omega_{n}\Arrowvert^{1}_{{\cal
    {U}}_{t_{1}}^{+}})^{2}\rightarrow 0$ which implies using the Sobolev embedding theorem that $(\Arrowvert
\omega_{n}\Arrowvert^{0}_{{\cal {U}}_{t_{1}}^{+}})^{2}\rightarrow 0$
hence $\eta =0$.

We define the functional $k_{f}(\square_{g}\omega)=\langle
f,\omega\rangle$. We show the functional is bounded so we can apply
Riesz's Theorem.  In that way $k_{f}$ define an element $\Psi \in
L^{2}({\cal {U}}_{t_{1}}^{+}, \nu_{g})$ such that
$k_{f}(\square_{g}\omega)=\langle\Psi,\square_{g}\omega\rangle$

Now using the energy inequality (\ref{v1}), \ref{v2}  and Cauchy-Schwartz we obtain:

\begin{eqnarray}
  k(\square_{g}\omega)&=&\langle f,\omega\rangle\\
      &\le&\langle f,f\rangle\langle\omega,\omega\rangle\\
      &\lesssim&(\Arrowvert f\Arrowvert^{0}_{{\cal {U}}_{t_{1}}^{+}})^{2}(\Arrowvert\square_{g}\omega\Arrowvert^{0}_{{\cal {U}}_{t_{1}}^{+}})^{2}
\end{eqnarray}

The Hahn-Banach theorem together with the fact that $\square_{g}
V_{\Sigma_{0}}$ is dense in $L^{2}({\cal {U}}_{t_{1}}^{+}, \nu_{g})$
allows us to extend the functional to the
whole $L^{2}({\cal {U}}_{t_{1}}^{+}, \nu_{g})$ without increasing the
norm.Hence, $k_{f}$ is bounded in $L^{2}({\cal {U}}_{t_{1}}^{+},
\nu_{g})$.

Then Riesz's Theorem implies there is a $\Psi\in L^{2}({\cal {U}}_{t_{1}}^{+}, \nu_{g})$ such that 
$$\int_{{\cal {U}}_{t_{1}}^{+}}\Psi\square_{g}\omega \nu_{g}=\int_{{\cal {U}}_{t_{1}}^{+}}f\omega \nu_{g}$$

for all $\omega\in V_{\Sigma_{t_{1}}}\supset C^{\infty}_{0}({\cal {U}}_{t_{1}}^{+})$. Then $\Psi$ is a solution of $\square_{g}\Psi=f$.

To add initial conditions we construct a specific function $q$ that satisfies the required initial conditions. We show this in detail in the following proposition:

\begin{Proposition}
There is a function $\phi$ in  $L^{2}({\cal {U}}_{t_{1}}^{+},\nu_{g})$  that satisfies:

\begin{itemize}
  \item  $\square_{g}\phi=f$
  \item  $\phi|_{\Sigma_{0}}=\varphi$  
  \item $\frac{\partial \phi}{\partial t}|_{\Sigma_{0}}=\pi$
\end{itemize}

with $\varphi,\pi \in H^{2}(\Sigma_{0},\mu_{h})\cap W^{1,2r}(\Sigma_{0},\mu_{h})$ where the connection is in $L^{2p}({\cal{U}}_{t_{1}}^{+},\nu_{g})$ and $\frac{1}{p}+\frac{1}{r}=1$ 
\end{Proposition}

{\it{Proof of Proposition 1}}. Let $q=\varphi+t\pi \in H^{1}({\cal{U}}_{t_{1}}^{+})$ which implies $q|_{\Sigma_{0}}=\varphi$ and  $\frac{\partial q}{\partial t}|_{\Sigma_{0}}=\pi$. Then 

 $$\square_{g} q=g^{ij}\frac{\partial q}{\partial x^{i}\partial x^{j}}+\Gamma^{k}_{ij}\frac{\partial q}{\partial x^{k}}.$$ 

 In order to that expression be in
 $L^{2}({\cal{U}}_{t_{1}}^{+},\nu_{g})$ we need to look for sufficient
 analytic conditions. The first term imposes that $\frac{\partial
   q}{\partial x^{i}\partial x^{j}}\in
 L^{2}({\cal{U}}_{t_{1}}^{+},\nu_{g})$ so $q$ must be in
 $H^{2}({\cal{U}}_{t_{1}}^{+},\nu_{g})$. To analyse the second term we
 use Holder's Inequality as follows:
  
 \begin{eqnarray}
    \int_{{\cal{U}}_{t_{1}}^{+}}(\Gamma^{k}_{ij}\frac{\partial q}{\partial x^{k}})^{2}\nu_{g}&\le&\left(\int_{{\cal{U}}_{t_{1}}^{+}}\left(\frac{\partial q}{\partial x^{k}}\right)^{2r}\nu_{g}\right)^{\frac{1}{r}} \left(\int_{{\cal{U}}_{t_{1}}^{+}}\left(\Gamma^{k}_{ij}\right)^{2p}\nu_{g}\right)^{\frac{1}{p}}\\
  \end{eqnarray}  
  
  where $\frac{1}{p}+\frac{1}{r}=1$. 
  
  If the connection only satisfies being in
  $L^{2}({\cal{U}}_{t_{1}}^{+},\nu_{g})$ then we need that $q\in
  W^{1,\infty}({\cal{U}}_{t_{1}}^{+})$. If on the other hand we know
  that the metric is $W^{1,\infty}({\cal{U}}_{t_{1}}^{+})$ then the
  condition $q\in H^{2}({\cal{U}}_{t_{1}}^{+},\nu_{g})$ can be
  maintained. Hence, sufficient conditions for the initial data are
  that $\varphi,\pi$ are in $H^{2}(\Sigma_{0},\mu_{h})\cap
  W^{1,2r}(\Sigma_{0},\mu_{h})$.

  Then applying Hahn-Banach theorem to the functional
  $k_{f-\square_{g} q}$ we obtain a function $\Psi'$ such
  that $$\int_{{\cal {U}}_{t_{1}}^{+}}\Psi'\square_{g}\omega
  \nu_{g}=\int_{{\cal {U}}_{t_{1}}^{+}}(f-\square_{g} q)\omega
  \nu_{g}$$ which again satisfies
  $\Psi'|_{\Sigma_{0}}=\frac{\partial\Psi'}{\partial
    t}|_{\Sigma_{0}}=0$

Then the desired solution is given by $\phi=\Psi'+q. \boxdot$\\

That there are solutions in $H^{k}({\cal{U}}_{t_{1}}^{+},\nu_{g})$ require higher energy inequality estimates. The main steps are to rewrite the wave equation as a first order system, obtain energy inequalities in Sobolev Spaces with negative integer and then apply standard dualities. (See \cite{hormander}, Chapter 23; \cite{ring}, Chapter 7).

\subsection{Uniqueness and continuity with respect initial data.}

The proof of uniqueness follows directly from (\ref{dasenergy}). Take two functions $\phi_{1}$ and $\phi_{2}$ in $H^{1}({\cal
  {U}}_{t_{1}}^{+},\nu_{g})$ such that solve the same initial value problem. Then we have that the
function $\tilde{\phi}=\phi_{1}-\phi_{2}$ satisfies (\ref{wp1}),
(\ref{wp2}) and (\ref{wp3}) with $f=0$ and vanishing initial
data. This implies:

\begin{eqnarray}
  (\Arrowvert \tilde{\phi}\Arrowvert^{1}_{{\cal {U}}_{t_{1}}^{+}})^{2}&\lesssim& \int^{t_{1}}_{0}\| \tilde{\phi}\|^{1}_{\Sigma_{\tau}})^{2} dt\\
  &\lesssim& \int^{t_{1}}_{0} (( \tilde{\|} \tilde{\phi}\|^{1}_{\Sigma_{0}})^{2} + (\Arrowvert f\Arrowvert_{U_{t}^{+}}^{0})^{2})dt\\
  &\lesssim& \int^{t_{1}}_{0} (( \tilde{\|} 0\|^{1}_{\Sigma_{0}})^{2} + (\Arrowvert 0\Arrowvert_{U_{t}^{+}}^{0})^{2})dt\\
 &=&0 
\end{eqnarray}

Because the norm of $(\Arrowvert \tilde{\phi}\Arrowvert^{1}_{{\cal {U}}_{t_{1}}^{+}})^{2}=0$ that implies $\tilde{\phi}=0$ and we have $$\phi_{1}=\phi_{2}.$$

In a similar way we prove the continuity of the solution with respect
initial data. We make the concept precise as follows. We say the
solution is continuously stable in $H^{1}({\cal
  {U}}_{t_{1}}^{+},\nu_{g})$ with respect initial data in $H^{2}({\cal
  {U}}_{t_{1}}^{+},\nu_{g})$ and source functions in $L^{2}({\cal
  {U}}_{t_{1}}^{+},\nu_{g})$ if for every $\epsilon'>0$ there is a
$\delta_{1},\delta_{2},\delta_{3}$ depending on $\varphi,\pi, f$ such
that if:

\begin{equation}
  (\Arrowvert \varphi-\tilde{\varphi}\Arrowvert^{2}_{\Sigma_{0}})^{2}\le \delta_{1},
  \end{equation} 

\begin{equation}
  (\Arrowvert \pi-\tilde{\pi}\Arrowvert^{2}_{\Sigma_{0}})^{2}\le \delta_{2}
  \end{equation} 

for $\varphi, \pi\in H^{2}(\Sigma_{0},\mu_{h})$ and
\begin{equation}
  (\Arrowvert f-\tilde{f}\Arrowvert^{0}_{{\cal{U}}_{t_{1}}^{+}})^{2}\le \delta_{3}
  \end{equation} 
for$f\in L_{2}({\cal {U}}_{t_{1}}^{+},\nu_{g})$ implies that 
\begin{equation}
  (\Arrowvert \phi-\tilde{\phi}\Arrowvert^{1}_{{\cal {U}}_{t_{1}}^{+}})^{2}\le \epsilon'
\end{equation} 
where $\tilde{\phi}$ is a solution with initial data given by
$\tilde{\phi}|_{\Sigma_{0}}=\tilde{\varphi}$ and
$\frac{\partial\tilde{\phi}}{\partial t}|_{\Sigma_{0}}=\tilde{\pi}$
and source function $\tilde{f}$.

Choose $$\delta_{1}=\frac{\epsilon}{3 t_{1}}$$

$$\delta_{2}=\frac{\epsilon}{3 t_{1}}$$

$$\delta_{2}=\frac{\epsilon}{3 t_{1}}$$
Then we obtain the inequalities:

\begin{equation}
  \sum_{\alpha\le1}\left(\int_{\Sigma_{0}}\left(\frac{\partial^{\alpha}\varphi}{\partial x^{\alpha}}\right)^{2 }-\left(\frac{\partial^{\alpha}\tilde{\varphi}}{\partial x^{\alpha}}\right)^{2 } \mu_{h}\right)\le \frac{\epsilon}{3 t_{1}}
\end{equation}

and 

\begin{equation}
\left(\int_{\Sigma_{0}}{\pi}^{2}-{\tilde{\pi}}^{2} \mu_{h}\right)\le \frac{\epsilon}{3 t_{1}}
\end{equation}
as a direct consequence of $(\Arrowvert \varphi-\tilde{\varphi}\Arrowvert^{2}_{\Sigma_{0}})^{2}\le \frac{\epsilon}{3 t_{1}}$ and $(\Arrowvert \pi-\tilde{\pi}\Arrowvert^{2}_{\Sigma_{0}})^{2}\le \frac{\epsilon}{3 t_{1}}$

Adding both inequalities we obtain:

\begin{equation}
  \sum_{\alpha\le 1}\left(\int_{\Sigma_{0}}\left(\frac{\partial^{\alpha}\varphi}{\partial x^{\alpha}}\right)^{2}-\left(\frac{\partial^{\alpha}\tilde{\varphi}}{\partial x^{\alpha}}\right)^{2} \mu_{h}\right)+\left(\int_{\Sigma_{0}}{\pi}^{2}-{\tilde{\pi}}^{2} \mu_{h}\right)\le \frac{2\epsilon}{3t_{1}}
\end{equation}

but this implies that:

\begin{equation}
  ( \tilde{\|}\phi-\tilde{\phi} \|^{1}_{\Sigma_{0}})^{2}\le\frac{2\epsilon}{3t_{1}}
\end{equation}

We also have that 

\begin{equation}
  (\Arrowvert f-\tilde{f}\Arrowvert^{0}_{{\cal{U}}_{t_{1}}^{+}})^{2}\le \frac{\epsilon}{ 3t_{1}}
  \end{equation} 

Then applying again the energy inequality we have:

\begin{eqnarray}
  (\Arrowvert \phi-\tilde{\phi}\Arrowvert^{1}_{{\cal {U}}_{t_{1}}^{+}})^{2}&\lesssim& \int^{t_{1}}_{0}\tilde{\|} \phi-\tilde{\phi}\|^{1}_{\Sigma_{t}})^{2} dt\\
  &\lesssim& \int^{t_{1}}_{0} (( \tilde{\|} \phi-\tilde{\phi}\|^{1}_{\Sigma_{0}})^{2} + (\Arrowvert f-\tilde{f}\Arrowvert_{U_{t}^{+}}^{0})^{2})dt\\
  &\lesssim& \int^{t_{1}}_{0}\frac{2\epsilon}{3t_{1}}+ \frac{\epsilon}{ 3t_{1}} dt\\
 &\lesssim& \epsilon\\
 &=&K\epsilon
\end{eqnarray}

Choosing $\epsilon=\frac{\epsilon '}{K}$ we obtain the required result.

\section{Discussion}\label{conlusion}

The equivalence between a solution of Einstein's Field Equations and the background metric on which the fields propagate makes the definition of a singularity in General Relativity subtler than in any other physical theory. The most common definition of a singular-free spacetime was formulated by Geroch \cite{geroch}; a spacetime is singular-free if it is geodesically complete. This characterisation is well motivated as we can associate the history of free-falling test particles to geodesic motion. In fact, Hawking and Penrose \cite{hawking} showed under very general physical conditions on the topology of the manifold and the behaviour of the causal structure of the spacetime that during gravitational collapse the spacetime can not be geodesically complete.  

 However, there is a gap between physical intuition and mathematical formalism. We want to describe from a physical point of view a singularity as some region of spacetime that becomes more and more pathological (maybe by the curvature becoming unbounded) where eventually General Relativity is no longer adequate and some Quantum Gravity theory is needed. The singularity theorems make a more modest conclusion simply that spacetime contains a geodesic that can not be continued indefinitely. The main objective of the present work is to show possible ways in which we can bridge the gap between intuitions and formalisms. 
   
In order to pursue a more complete physical description of singularities we proposed to define the notion of a singularity as obstruction to fields rather that geodesic trajectories. Then one can talk about the field regularity of spacetime.  In particular, in this paper the concept of classical strongly wave regularity was explored on spacetimes with low differentiability with certain timelike vector field transverse to caustics or shocks where the curvature is not a continuous function. The paper showed that the wave equation has unique solutions on the Sobolev space $H^{1}({\cal{U}}_{t_{1}}^{+},\nu_{g})$ and that solutions depends continuously with respect initial data in $H^{2}(\Sigma_{0},\mu_{h})\times H^{2}(\Sigma_{0},\mu_{h})$.

In this sense the classical singularity does not disrupt the wave dynamics. We will further explore the field regularity of these spacetimes by extending this work by linking it with quantum field theory. That means to study the quantum wave regularity of the spacetime. Moreover, a concept of singular behaviour defined in terms of quantum fields rather than geodesics will provide a semi-classical picture able to use concepts from Quantum Field Theory and General Relativity. This type of analysis can then shed light on the behaviour we might find in quantum gravity.

In addition, the notion of well-posedness is related
  to the determinism of the field equations, and in the context of
  General Relativity this is intuitively the content of the Strong Cosmic
  Censorship Conjecture. Current approaches to giving a precise formulation of 
  this difficult conjecture rely precisely on the degree of differentiabilty and the 
  notion of maximal globally hyperbolic
  development \cite{Dafermos}. It would be interesting to reformulate the conjecture
  in terms of field regularity, which can be seen as a generalisation
  of the globally hyperbolic development in the context of low
  differentiability.  This will perhaps allow one to look at the conjecture in a
  new way.

\section*{Acknowledgements}

The author would like to thank J. Vickers for discussion and comments on and early draft. This work was supported by a CONACyT Graduate Fellowship.

\section{Appendix}\label{Appendix}

In the next Proposition we show that the there is a congruence of timelike geodesics whose tangent vector has an essentially bounded weak derivative. This is a key requirement in the above discussion to make sense of the energy inequalities as in a general low differentiable spacetime the covariant derivative may be unbounded and the argument breaks down.

\begin{Proposition}\label{append}
Let ($M,g$) be a Lorentzian manifold and $p$ a point in an open subset $\Omega\subset M$ with compact closure such that there is a lens-shaped domain ${\cal{U}}$ satisfying the condition of Lemma \ref{main}. 

Then there exists a congruence of timelike geodesics whose tangent vector, $\Upsilon^{a}$, has an essentially bounded weak covariant derivative.
\end{Proposition}

{\bf{Summary of the proof of Lemma. }}

The proof will consist of eight steps. 
\begin{enumerate}
  \item The first step defines the geometric setting and the specific class of mollifiers that are used.
  \item The second step focuses establishing a majorizing ordinary differential equation that helps to uniformly bound the norm of the tangent vectors of the mollified geodesics.
  \item The third step establishes a uniform time for existence of the mollified geodesics such that the tangent vectors of the mollified geodesics, $\gamma_{n}$ are contained inside the set $C$ (see hypothesis 5 of Lemma (\ref{main})). 
  \item The fourth step focuses on establishing a majorizing ordinary differential equation that helps to essentially bound the geodesic connecting vector $Y$.
  \item The fifth step establishes that $Y$ is essentially bounded.
  \item The sixth step uses the Arzela-Ascoli theorem and the bounds previously obtained to show that the limit $\lim_{n\rightarrow\infty}\gamma_{n}$ is well-defined and gives meaning to the notion of geodesic and tangent vector.
  \item The seventh step establishes the distributional nature of the weak covariant derivative of the tangent vector.
  \item The eighth step shows the essential boundedness of the weak covariant derivative.
\end{enumerate}

Remark: Using the specific class of mollifiers described below in step 1 is important in step 4 where it allows us to establish the conjecture of Clarke.

{\it{First Step.}}

Let ${\cal{U}}$ be a lens-shaped domain contained in compact set $\Omega$ that contains $p$. Also assume there is an appropriate choice of coordinates $x^{i}=\{\tau,x^{\alpha}\}$ on ${\cal{U}}_{t_{1}}^{+}$ such that $0\le\tau\le t_{1}$ (where $t_{1}$ is defined in the second step) and the constant values of $\tau$ correspond to the spacelike surfaces $\Sigma_{\tau}$, $\Sigma=\Sigma_{0}$ . (see figure 1)
  
  Now we choose a vector $\vec{V}\in C\subset\mathbb{R}^{4}$. Then define the vector $\vec{V_{q}}$ as the vector in $q\in \Sigma$ whose components in the vector basis of the coordinates $\{\tau,x^{\alpha}\}$ are the same as the components of $\vec{V}$ in the canonical basis in $\mathbb{R}^{4}$.
  
   The hypothesis of the theorem only allow to define the connection as a distribution. In order to use the differential equations in the classical sense we will use convolutions and then take limits. 
   
   The  strict delta net $(\rho_{n})_{n}$ that we use is defined using the admissible mollifiers described by Steinbauer and Vickers in \cite{gerocht}. We describe them below: 
   
   \begin{Definition}
  There exist strict delta net $(\rho_{n})_{n}$  with
  \begin{itemize}
      \item $supp(\rho_{n})\subset B_{\frac{1}{n}}(0) \text{ for all } n\in\mathbb{N}$
    \item $\int_{\mathbb{R}^{4}}\rho_{n}=1 \text{ for all } n\in\mathbb{N}$
    which are moderate, have vanishing moments and the negative parts have arbitrarilly small $L^{1}(\mathbb{R}^{4},d^{4}x)$-norm. This last condition means that
    \item $\forall \eta>0 \, \exists N_{0}:\int_{\mathbb{R}^{4}}|\rho_{n}|=1+\eta \text{ for all } n>N_{0}$
  \end{itemize} 
   \end{Definition}

    These mollifiers have the properties that if the components of $g_{ab}$ are in $H_{loc}^{1}({M},\nu_{g})\cap L_{loc}^{\infty}(M,\nu_{g})$ then the convolution of the components $g_{ab}$ with the admissible mollifiers denoted by $( g_{ab})_{n}=g_{ab}*\rho_{n}$ satisfy:
   \begin{itemize}
     \item $(g_{ab})_{n}\rightarrow g_{ab}$ in $H_{loc}^{1}({M},\nu_{g})\cap L_{loc}^{\infty}(M)$
     \item $((g_{ab})_{n})^{-1}\rightarrow g^{ab}$ in $H_{loc}^{1}({M},\nu_{g})\cap L_{loc}^{p}(M,\nu_{g})$ for all $p<\infty$
     \item $\Gamma_{ijk}((g_{ab})_{n})\rightarrow \Gamma_{ijk}$ in $L_{loc}^{2}({M},\nu_{g})$
     \item $R^{i}_{jkl}((g_{ab})_{n})\rightarrow R^{i}_{jkl}$ in $D'({\cal{U}}_{t_{1}}^{+})$
   \end{itemize}
    where $\Gamma_{ijk}((g_{ab})_{n})$ is the Christoffel symbols of the metric $( g_{ab})_{n}$ and $R^{i}_{jkl}((g_{ab})_{n})$ is the curvature of the metric $( g_{ab})_{n}$

For convenience we are going to denote $\Gamma_{ijk}((g_{ab})_{n})=\Gamma_{ijk (n)}$ and in general all Christoffel symbols appearing will be understand as those coming from the convolution of the metric by the corresponding element of the delta net. \\
 
{\it{Second Step.}}

  Now we define for each point $q$ in $\Sigma$, a family of geodesics $\{\gamma^{q}_{n}(s)\}$that satisfies:
  
  \begin{equation}\label{ge}
    \frac{d^{2}\gamma^{q}_{n}}{ds^{2}}^{i}=-{\Gamma}_{jk}^{i (n)}\frac{d\gamma^{q}_{n}}{ds}^{j}\frac{d\gamma^{q}_{n}}{ds}^{k}
  \end{equation}
    
    with the initial conditions 
    
    \begin{eqnarray}
       \gamma^{q}_{n}(0)=q\\
       \frac{d\gamma^{q}_{n}}{ds}(0)=\vec{V_{q}}
 \end{eqnarray}
   
We now show that there is some time $s_1$ (uniform in $n$) such that for $|s| < s_1$ we have  $\frac{d\gamma_{n}}{ds}(s)\in C$. We will drop the $q$ in the notation until it is needed.

  First we define: 
   
\begin{equation}\label{r}
  l=sup\{\Arrowvert\vec{V}\Arrowvert,:\vec{V}\in C  \}
  \end{equation}
  \begin{equation}\label{l}
     r=inf\{dist(\frac{d\gamma_{n}}{ds}(0),C^{c}):q\in \Sigma\} 
\end{equation} 
where \emph{dist} means the euclidean distance and where $C^{c}$ is the complement of $C$. We first give a uniform bound $\left\Arrowvert\frac{d\gamma_{n}}{ds}\right\Arrowvert$ in terms of $s$ and the intial value. To do this we note that: 

\begin{eqnarray}
\left(\left\Arrowvert\frac{d\gamma_{n}}{ds}\right\Arrowvert\right)\frac{d}{ds}\left\Arrowvert \frac{d\gamma_{n}}{ds} \right\Arrowvert&\le&\frac{1}{2} \frac{d}{ds}\left\Arrowvert\frac{d\gamma_{n}}{ds}\right\Arrowvert^{2}\\
&=&\frac{d}{ds}\left(\frac{d\gamma_{n}}{ds}\right)\cdotp \frac{d\gamma_{n}}{ds}\\
&\le&\left\Arrowvert\frac{d}{ds}\left(\frac{d\gamma_{n}}{ds}\right)\right\Arrowvert \left\Arrowvert \frac{d\gamma_{n}}{ds} \right\Arrowvert
\end{eqnarray}

where $u\cdotp v$ is the dot product in $\mathbb{R}^{4}$.

Hence

\begin{equation}\label{fammain}
  \frac{d}{ds}\left\Arrowvert \frac{d\gamma_{n}}{ds} \right\Arrowvert\le \left\Arrowvert\frac{d}{ds}\left(\frac{d\gamma_{n}}{ds}\right)\right\Arrowvert
\end{equation}

Consider now the following inequalities:

\begin{eqnarray}
  \left\Arrowvert\frac{d}{ds}\left(\frac{d\gamma_{n}}{ds}\right)\right\Arrowvert&\le&2 sup_{i=0,..,3}|\left\arrowvert\frac{d}{ds}\left(\frac{d\gamma^{i}_{n}}{ds}\right)\right\arrowvert\\
  &=&sup_{i=0,..,3}\left\arrowvert-{\Gamma}_{jk}^{i (n)}\frac{d\gamma^{q}_{n}}{ds}^{j}\frac{d\gamma^{q}_{n}}{ds}^{k}\right\arrowvert\\
  &\le&32 sup_{i,j,k}\left\arrowvert-{\Gamma}_{jk}^{i (n)}\right\arrowvert sup_{j,k=0,..,3}\left\arrowvert\frac{d\gamma^{q}_{n}}{ds}^{j}\frac{d\gamma^{q}_{n}}{ds}^{k}\right\arrowvert\\
  &\le& 32 sup_{i,j,k}\left\arrowvert-{\Gamma}_{jk}^{i (n)}\right\arrowvert \left\Arrowvert\frac{d\gamma_{n}}{ds}\right\Arrowvert^{2}
  \end{eqnarray}

which using (\ref{fammain}) gives:

\begin{equation}
  \frac{d}{ds}\left\Arrowvert \frac{d\gamma_{n}}{ds} \right\Arrowvert\le 32 sup_{i,j,k}\left\arrowvert-{\Gamma}_{jk}^{i (n)}\right\arrowvert \left\Arrowvert\frac{d\gamma_{n}}{ds}\right\Arrowvert^{2}
\end{equation}

So  $\left\Arrowvert\frac{d\gamma_{n}}{ds}\right\Arrowvert$ is bounded by the majorizing equation 

\begin{equation}
  \frac{dx}{ds}=\lambda^{n} x^{2}
\end{equation}

subject to the initial condition $x(0)<l$ and where $\lambda^{n}(s):=32 sup_{i,j,k}\left\arrowvert-{\Gamma}_{jk}^{i (n)}\right\arrowvert$

Then we have:

\begin{eqnarray}
   && \frac{dx}{ds}\le\lambda^{n} x^{2}\\
    &\Rightarrow& -\frac{d}{ds}\left(\frac{1}{x}\right)\le\lambda^{n}(s)\\
    &\Rightarrow& \frac{1}{x(s)}-\frac{1}{x(0)}\le-\int^{s}_{0}\lambda^{n}(s') ds'\\
    &\Rightarrow& x(s)<\frac{x(0)}{1-x(0)\int^{s}_{0}\lambda^{n}(s') ds'}<k x(0)\label{majo}
\end{eqnarray}

for $k>\frac{1}{1-x(0)\int^{s}_{0}\lambda^{n}(s') ds'}>1$ \\

\emph{Third Step.}

We now use the result above to show the existence of a time interval for which $\dot \gamma$ remains in $C$. 

Notice that
\begin{eqnarray}
  \int^{s}_{0}\arrowvert\Gamma_{jk}^{i(n)}\arrowvert ds &=& \int^{s}_{0}\arrowvert \rho_{n}\star\Gamma_{jk}^{i} \arrowvert ds'\\
                                                     &=&\int^{s}_{0}\arrowvert \int_{\mathbb{R}^{4}} \rho_{n}(z)\Gamma_{jk}^{i}(\gamma(s')+z)dz\arrowvert ds' \\
                                                     &\le&\int^{s}_{0} \int_{\mathbb{R}^{4}} \arrowvert \rho_{n}(z)\arrowvert\arrowvert\Gamma_{jk}^{i}(\gamma(s')+z)\arrowvert dzds' \\
                                                       &\le&  \int_{\mathbb{R}^{4}} \arrowvert \rho_{n}(z)\arrowvert\int^{s}_{0}\arrowvert\Gamma_{jk}^{i}(\gamma(s')+z)\arrowvert ds'dz\\ 
                                                       &\le& M(s)\int_{\mathbb{R}^{4}}\arrowvert \rho_{n}(z)\arrowvert dz \\
                                                      &\le& M(s)(1+\eta) \label{s0}
   \end{eqnarray}
 
for $ \eta>0$ when $n\ge  N_{0}$ where we have use (\ref{majo}) with $k=1+\frac{r}{x(0)}$ so that $\dot{\gamma}\in C$.

Then $g^{n}(s)=\int^{s}_{0}\lambda^{n}(s') ds'$ is a non decreasing
function that satisfies $\lim_{s^{+}\rightarrow 0} g^{n}(s)=0$ for all
$n$ as a consequence of property (5) of the hypothesis of Lemma
(\ref{main}) and (\ref{s0}). Moreover, there is a smallest time
$s_{0}\neq 0$ such that
$k=1+\frac{r}{x(0)}=\frac{1}{1-x(0) g^{n_{0}}(s_{0})}$ for some $n_0$.
Then choosing $s < s_{0}$ guarantees that
$1+\frac{r}{x(0)}>\frac{1}{1-l\int^{s}_{0}\lambda^{n}(s') ds'}$ for
all $n$. Notice that $s_{0}$ can not be $0$ because that would imply $k=1$
contradicting $k>1$.

 Then for $s<s_{a}$
integrating and taking absolute values on the geodesic equation
(\ref{ge}) we have that
\begin{eqnarray}
  \arrowvert\frac{d\gamma_{n}}{ds}^{i}(s)-\frac{d\gamma_{n}}{ds}^{i}(0)\arrowvert &\le&\int^{s}_{0}\arrowvert{\Gamma}_{jk}^{i(n)}\frac{d\gamma_{n}}{ds}^{j}\frac{d\gamma_{n}}{ds}^{k}\arrowvert ds'\\
  &\le& 16 (r+l)^{2}\int^{s}_{0}\arrowvert{\Gamma}_{jk}^{i(n)}\arrowvert ds'\\
  &\le& 16 (r+l)^{2}M(s)(1+\eta)
    \end{eqnarray}

Now because $M(s)\rightarrow 0$ as $s\rightarrow 0$ we can make the difference as small as we want. So taking explicitly $s_{1}\le s_{0}$ such that
\begin{equation}
  M(s_{1})<\frac{r}{32(r+l)^{2}(1+\eta)}
\end{equation}

Ensures that:

\begin{eqnarray}\label{comp}
  \left\Arrowvert\frac{d\gamma_{n}}{ds}(s)-\frac{d\gamma_{n}}{ds}(0)\right\Arrowvert&\le&2 sup_{i=0,..,3}\left\arrowvert\frac{d\gamma_{n}}{ds}^{i}(s)-\frac{d\gamma_{n}}{ds}^{i}(0)\right\arrowvert \\\label{s1}
                                                                                         &\le& 32 (r+l)^{2}M(s)(1+\eta)\\\label{s2}
 &\le& r
\end{eqnarray}                                                                                                                                                                        

for all $s\le s_{1}$ (where $s_{1}$ is independent of $n$). This implies that $\frac{d\gamma_{n}}{ds}(s)\in C$. Then we can choos $t_{1}$ as sufficiently close to $0$ to ensure that $p$ is covered by the curves up to $s_{1}$\\

{\it{Fourth Step.}}

We now consider a $1$-parameter family of initial conditions 

 \begin{eqnarray}
       \gamma^{q}_{n}(0,u)=q(u)\\
       \frac{d\gamma^{q}_{n}}{ds}(0,u)=\vec{V}_{q(u)}
 \end{eqnarray}

and the corresponding family of geodesics, $\{\gamma^{q(u)}_{n}(s)\}_{u}$. Now let $Y$ be the connecting vector of this family. For simplicity later on, we are going to use a parallel propagated co-frame on $\gamma^{q}_{n},\{e^{a (n)}=e^{a (n)}_{i}dx^{i}\}_{a=0,1,2,3}$  coinciding with the coordinate basis at $s=0$ so we define:

\begin{equation}
  Y_{u}^{a}:=\frac{\partial \gamma^{q}_{n}}{\partial u}^{i}e^{a (n)}_{i}:=Y_{u}^{i}e^{a (n)}_{i}
\end{equation}

Moreover if $q(u)$ is a coordinate function of $\Sigma$ i.e., $q(u)=x^{\alpha}$, we will denote the connecting vector as $Y_{\alpha}$ with frame components $Y_{\alpha}^{a}$.
Now we have

\begin{eqnarray}
  \frac{D^{2} Y_{\alpha}^{a}}{Ds^{2}}&=&\frac{D}{Ds}\left(\frac{dY_{\alpha}^{a}}{ds}\frac{\partial}{\partial s}\right)\\
                                      &=&\frac{d^{2}Y_{\alpha}^{a}}{ds^{2}}\frac{\partial}{\partial s}\label{y2}
\end{eqnarray}
where we have use the fact that $Y_{\alpha}^{a}$ is a scalar and $\frac{D}{Ds}\left(\frac{\partial}{\partial s}\right)$ is the geodesic equation where we have define $\frac{D}{Ds}:=\nabla_{\frac{\partial}{\partial s}}$.

Now notice that in frame components we have

\begin{eqnarray}
  \frac{D^{2} Y_{\alpha}^{a}}{Ds^{2}}&=&\frac{D}{Ds}\left(\frac{DY_{\alpha}^{i}}{Ds}e^{a (n)}_{i}+ Y_{\alpha}^{i}\frac{De^{a (n)}_{i}}{Ds}\right)\\
                                     &=&\frac{D}{Ds}\left(\frac{dY_{\alpha}^{i}}{ds}e^{a (n)}_{i}\frac{\partial}{\partial s}+ Y_{\alpha}^{i}\frac{De^{a (n)}_{i}}{Ds}\right)\\ 
                                                                          &=&\frac{D^{2}Y_{\alpha}^{i}}{Ds^{2}}e^{a (n)}_{i}\frac{\partial}{\partial s}\label{y1}
                                                         \end{eqnarray}
where we have use the fact that, $Y_{\alpha}^{a}$ is a scalar and $e^{a (n)}_{i}$ are parallel propagated coefficients.

Now using the geodesic deviation equation and using (\ref{y1}) and (\ref{y2}) we have:

\begin{eqnarray}
  \frac{d^{2}Y_{\alpha}^{a}}{ds^{2}}&=&\frac{D^{2}Y_{\alpha}^{a}}{Ds^{2}}\\
                                     &=& \frac{D^{2}Y_{\alpha}^{i}}{Ds^{2}}e^{a (n)}_{i}\\
                                     &=& e^{a (n)}_{i}R^{i}_{jkl}\frac{d\gamma_{n}}{ds}^{j}\frac{d\gamma_{n}}{ds}^{k}Y_{\alpha}^{b}e_{b }^{l(n)}\label{gd}
\end{eqnarray}
  
with the initial conditions:

\begin{eqnarray}
    Y_{\alpha}^{a}(0)&=&Y^{i}_{\alpha}(0)e^{a (n)}_{i}(0)\\
                    &=&\delta^{i}_{\alpha}\delta^{a}_{i}\\
                    &=&\delta^{a}_{\alpha}
\end{eqnarray}

where the first $\delta$ is by noting that at $s=0$ we have $\frac{\partial \gamma^{q i}_{n}}{\partial x^{\alpha}}=\frac{\partial (x^{i})^{q }_{n}}{\partial x^{\alpha}}$ and the second one as a consequence of the initial alignment with the coordinates.

Also notice that

\begin{eqnarray}
  \frac{dY_{\alpha}^{a}}{ds}e^{(n)}_{a}&=&\left(\nabla_{\frac{\partial}{\partial s}}Y_{\alpha}^{a}\right)e^{(n)}_{a}\\
                                &=& \left(\nabla_{\frac{\partial}{\partial s}}Y_{\alpha}^{a}\right)e^{(n)}_{a}+Y_{\alpha}^{a}\nabla_{\frac{\partial}{\partial s}}e^{(n)}_{a}\\
                                &=&\nabla_{\frac{\partial}{\partial s}}Y_{\alpha}^{a}e^{(n)}_{a}\\\label{ic}
                                &=&\nabla_{Y_{\alpha}^{a}e^{(n)}_{a}}\frac{\partial}{\partial s}\\
                                &=&Y_{\alpha}^{a}\nabla_{e^{(n)}_{a}}\frac{\partial}{\partial s}
                                                \end{eqnarray}
where (\ref{ic}) is the use of the torsion free condition. Now if we evaluate at $s=0$ we have the second initial condition by the following calculation 

\begin{eqnarray}
  \frac{dY_{\alpha}^{a}}{ds}(0)e^{(n)}_{a}&=&Y_{\alpha}^{a}(0)\nabla_{e^{(n)}_{a}}\frac{\partial}{\partial s}|_{0}\\
                                    &=&\delta_{\alpha}^{a}\nabla_{e^{(n)}_{a}(0)}\frac{\partial}{\partial s}|_{0}\\
                                    &=&\nabla_{\frac{\partial}{\partial x^{\alpha}}}\frac{\partial}{\partial s}|_{0}\\
                                    &=&\nabla_{\frac{\partial}{\partial x^{\alpha}}}\vec{V}_{q}\\
                                    &:=&\vartheta \label{defii}
                                                                     \end{eqnarray}
                                
where (\ref{defii}) is a definition.

Now notice that:

\begin{equation}
\nabla_{\frac{\partial}{\partial s}}e^{a (n)}=0  
\end{equation}
  
  as a consequence of being parallel propagated.
  
   Then in coordinates we have:

  \begin{equation}\label{e}
  \frac{de^{a (n)}_{i}}{ds}=\Gamma_{ij}^{k (n)}e_{k}^{a (n)}\frac{d\gamma_{n}}{ds}^{j}
  \end{equation}

So using again a majorizing equation $$\frac{dy}{ds}=c\lambda^{n}y$$ with initial condition $z(0)=1$, $c$ a constant and applying similar arguments as the ones used in step two and three we can conclude that $e^{a (n)}_{i}$ is uniformly bounded in terms of $M(s_{1})$.

It follows then that using (\ref{gd}) and the uniform bounds on $e^{a (n)}_{i}$ and $\frac{d\gamma_{n}}{ds}$ that we can write:

\begin{equation}
   |\frac{d^{2}Y_{\alpha}^{a}}{ds^{2}}|\le C\sigma\|Y_{\alpha}^{a}\|_\infty
  \end{equation} 
                                
where C is a suitable constant and
\begin{equation}
  \sigma:=sup |R^{(n)i}_{jkl}(\gamma_{n}(s))|
\end{equation}

Next we consider the majorizing equation

\begin{equation}\label{z}
  \frac{d^{2}z}{ds^{2}}=C\sigma z
\end{equation}

with initial conditions

\begin{equation}
 z(0)=1,\ \ \frac{dz(0)}{ds}=sup|  \vartheta|
\end{equation}

where the supremum is taken with respect all indices that appear in $\vartheta$.

Notice that bounding $z$, implies a bound of $\|Y_{\alpha}^{a}\|_\infty$. \\

{\it{Fifth Step.}}

We now obtain a bound on $z$ in terms of the curvature tensor. For that purpose we need the following inequality:

\begin{eqnarray}\label{sig}
  \int^{s}_{0}|R^{(n)i}_{jkl}|ds&\le&\int^{s}_{0}\sigma ds'\\
  &=& \int^{s}_{0}sup\arrowvert \rho_{n}\star R_{jkl}^{i} \arrowvert ds'\\
                                                     &=&\int^{s}_{0}sup\arrowvert \int \rho_{n}(z)R_{jkl}^{i}(\gamma(s')+z)dz\arrowvert ds' \\                                               
                                                       &\le&  \int \arrowvert \rho_{n}(z)\arrowvert\int^{s}_{0}sup\arrowvert R_{jkl}^{i}(\gamma(s')+z)\arrowvert ds'dz\\ 
                                                       &\le& N(s)\int\arrowvert \rho_{n}(z)\arrowvert dz \\
                                                       &\le& N(s)
   \end{eqnarray}

        To estimate a solution to (\ref{z}) we notice that if $s_{2}$ is the first value of $\Sigma$ at which $|z|=2$ (possibly $s_{2}=\infty$) then before $s_{2}$
        
        \begin{eqnarray}
          \frac{d^{2}z}{ds^{2}}&=&C \sigma z\\
                               &\le&  C \sigma 2 
        \end{eqnarray} 
     which is a consequence of the initial conditions and the continuity of $z$.
     
   Now integrating two times both sides considering (\ref{sig}) and the initial conditions on $z$ we got
   
   \begin{equation}
     z\le 1+sup| \vartheta|s +2C\int^{s} N(s')ds'
   \end{equation}
   
   for $0\le s\le s_{2}\le s_{1}$
   
   Now the right side is an increasing function that starts at  zero. So there is a $s_{3}$ such that
   
   \begin{equation}
     | \vartheta|s_{3} +2C\int^{s_{3}}N(s')ds'\le1
   \end{equation}
     then we will have $z\le2$ up to $s_{3}$, and hence 
     \begin{equation}\label{boundy}
      \|Y_{\alpha}^{a}\|_\infty\le2 
     \end{equation}
   in this interval.\\
     
     {\it{Sixth Step.}}
     
     The Arzela-Ascoli theorem guarantees that given $\{\rho_{n}\}_{n}$,a sequence of equicontinuous and uniformly bounded functions there is a sub-sequence that converges uniformly. 
    
    Now the functions $$\{\gamma^{n}:(q,s)\rightarrow \gamma^{q}_{n}(s)\}$$ are equicontinuous and uniformly bounded. This can be seen by noting that the functions are defined in a bounded domain with a direct application of the mean value theorem in several variables and the fact that $\frac{\partial\gamma_{n}^{q }}{\partial x^{j}}$ and $\frac{\partial\gamma_{n}^{q }}{ds}$ are uniformly bounded.   
        
         So by Arzela -Ascoli theorem there is a sub-sequence of $\{\gamma^{n}\}$ that gives meaning to the idea of a geodesic $\gamma $ with tangent vector $\Upsilon$. \\

       {\it{Seventh Step.}}

        We now define the $j$ component of $\left(\nabla_{i}\frac{d\gamma_{n}}{ds}\right)$ as:
        \begin{eqnarray}\label{cdt}
                  \left(\left(\nabla_{i}\frac{d\gamma_{n}}{ds}\right)^{j},\phi\right)                  &=&\int_{{\cal{U}}_{t_{1}}^{+}}\left(\nabla_{i}\frac{d\gamma_{n}}{ds}\right)^{j} \phi\nu_{g}\\
                    &=&\int_{{\cal{U}}_{t_{1}}^{+}}\left(\frac{\partial}{\partial x^{i}}\left(\frac{d\gamma_{n}}{ds}^{j}\right)+\Gamma_{ki}^{(n) j}\frac{d\gamma_{n}}{ds}^{k}\right) \phi \nu_{g}\\
                    &=&-\int_{{\cal{U}}_{t_{1}}^{+}}\frac{d\gamma_{n}}{ds}^{k}\left(\frac{\partial\phi}{\partial x^{i}}\delta_{k}^{j}-\Gamma_{ki}^{(n) j}\phi\right) \nu_{g}\label{di}
        \end{eqnarray}
        
 where (\ref{di}) is obtained by integrations by parts.
 
  Notice that the right hand converges in $\mathbb{R}$ as $n$ tends to infinity for every $\phi\in {\cal{D}}({{{\cal{U}}_{t_{1}}^{+}}})$. Hence, the expression converges in the sense of distributions to the distributional covariant derivative of  $\Upsilon $ (see \cite{pde}, p. 134).\\

{\it{Eight Step.}}

  We now establish the essential boundedness of the weak covariant derivative of the tangent vector. 
 Consider a basis for $T_{p}{\cal{U}}_{t_{1}}^{+}$ is $\{\frac{d\gamma_{n}}{ds}, Y_{\alpha}^{i}\frac{\partial}{\partial x^{i}}\}_{\alpha={1,2,3}}$. So any vector $X$ can be written as a linear combination of those.
 
If the $i$- component of $X$ is $X^{i}=X^{\alpha}Y_{\alpha}^{i}+X^{0}\frac{d\gamma_{n}}{ds}^{i}$ then

\begin{eqnarray}
  \nabla_{X}\frac{d\gamma_{n}}{ds}&=&\nabla_{X^{i}\frac{\partial}{\partial x^{i}}}\frac{d\gamma_{n}}{ds}\\
                                  &=&\nabla_{X^{\alpha}Y_{\alpha}^{i}\frac{\partial}{\partial x^{i}}}\frac{d\gamma_{n}}{ds}+\nabla_{X^{0}\frac{d\gamma_{n}}{ds}^{i}\frac{\partial}{\partial x^{i}}}\frac{d\gamma_{n}}{ds}\\
                                  &=&X^{\alpha}\nabla_{Y_{\alpha}}\frac{d\gamma_{n}}{ds}+X^{0}\nabla_{\frac{d\gamma_{n}}{ds}}\frac{d\gamma_{n}}{ds}\\
                                  &=&X^{\alpha}\nabla_{\frac{d\gamma_{n}}{ds}}Y_{\alpha}\\
                                  &=&X^{\alpha}\nabla_{\frac{d\gamma_{n}}{ds}}Y^{a}_{\alpha}e^{(n)}_{a}\\
                                  &=&X^{\alpha}\frac{dY^{a}_{\alpha}}{ds}e^{(n)}_{a}
 \end{eqnarray}       
          where we have used the linearity of the covariant derivative, the torsion free condition and the vanishing Lie bracket between $Y_{\alpha}, \mbox{and} \frac{d\gamma_{n}}{ds}$ along with the geodesic equation.
          
          Now integrating once (\ref{gd}) we have:
          \begin{equation}
            \frac{dY^{a}_{\alpha}}{ds}=\frac{d\gamma_{n}^{ k}}{ds}(0)  \Gamma^{a (n)}_{\alpha k}(\gamma_{n}(0)) +\int_{0}^{s}  e^{a (n)}_{i}R^{i}_{jkl}\frac{d\gamma_{n}}{ds}^{j}\frac{d\gamma_{n}}{ds}^{k}Y_{\alpha}^{b}e_{b }^{l(n)} ds'
          \end{equation}
          
        Now this implies that $\left(\nabla_{i}\frac{d\gamma_{n}}{ds}\right)^{j}$ is bounded and the bound is independent of $n$. This can be seen by using (\ref{comp}), (\ref{e}),(\ref{boundy}) to bound $\frac{dY^{a}_{\alpha}}{ds}$ and the fact that $X^{\alpha}$ are continuous functions (any vector $X$ is a tangent vector of a $C^{1}$ curve) in a compact set so a bound exist.

           Now using the boundedness of $\left(\nabla_{i}\frac{d\gamma_{n}}{ds}\right)^{j}$ and (\ref{cdt}) we have an estimate of the form:
           
           \begin{equation}\label{eb}
             \left(\left(\nabla_{i}\frac{d\gamma_{n}}{ds}\right)^{j},\phi\right)\le B\| \phi\| _{1}
           \end{equation}
           
           where $B$ is a constant and $\| f\| _{1}=\int f$.

Moreover, (\ref{eb}) allows us to define $\left(\nabla_{i}\frac{d\gamma_{n}}{ds}\right)^{j}$ as a functional over the space of integrable functions, $L^{1}({\cal{U}}_{t_{1}}^{+},\mu_{g})$.
Now taking the limit as $n\rightarrow \infty$ we obtain
 
  \begin{equation}\label{e3b}
          \lim_{n\rightarrow\infty}  \left(\left(\nabla_{i}\frac{d\gamma_{n}}{ds}\right)^{j},\phi\right)\le B\| \phi\| _{1}
           \end{equation}
because all the bounds hold in the limit.

So $\lim_{n\rightarrow\infty} \left(\nabla_{i}\frac{d\gamma_{n}}{ds}\right)^{j}$ converge in the dual space (the space of linear functionals) of integrable functions.
 This space is isomorphic to  $L^{\infty}({\cal{U}}_{t_{1}}^{+})$ because ${\cal{U}}_{t_{1}}^{+}$ is compact. Then  $\left(\nabla_{i}\frac{d\gamma_{n}}{ds}\right)^{j}\in L^{\infty}({\cal{U}}_{t_{1}}^{+})$ and is a uniformly bounded function. 
Finally we have that $\lim_{n\rightarrow\infty} \nabla_{i}\frac{d\gamma_{n}}{ds}$ is essentially bounded because each component is an essentially bounded function.

\end{document}